\documentclass[epj]{svjour}

\usepackage{amsfonts}   
\usepackage{amsmath}  
\usepackage{color}
\usepackage{dcolumn}
\usepackage{bm}
\usepackage{graphicx}
\usepackage{epsfig}
\usepackage{color}
\usepackage{nicefrac}
\usepackage{transparent}

\newcommand{\bolde}{\boldsymbol{\varepsilon}}
\newcommand{\ua}{\uparrow}
\newcommand{\da}{\downarrow} 
\newcommand{\be}{\begin{equation}}
\newcommand{\ee}{\end{equation}}
\newcommand{\bea}{\begin{eqnarray}}
\newcommand{\ba}{\begin{array}}
\newcommand{\eea}{\end{eqnarray}}
\newcommand{\ea}{\end{array}}

\newcommand{\black}{black}
\newcommand{\tc}[1]{\textcolor{\black}{#1}}

\begin{document}

\title{\tc{Fisher information of Markovian decay modes}}
\subtitle{\tc{Nonequilibrium equivalence principle, dynamical phase transitions and coarse graining}}

\author{Matteo Polettini \inst{1}
\thanks{\emph{Present address:} matteo.polettini@uni.lu}
}  

\institute{Complex Systems and Statistical Mechanics, University of Luxembourg,\\ Campus Limpertsberg, 162a avenue de la Fa\"iencerie, L-1511 Luxembourg (G. D. Luxembourg)}

\date{}

\abstract{
\tc{We introduce the Fisher information in the basis of decay modes of Markovian dynamics, arguing that it encodes important information about the behavior of nonequilibrium systems. In particular we generalize} an orthonormality relation between decay eigenmodes of \tc{detailed balanced} systems to  \tc{normal} generators that commute with their time-reversal. Viewing such modes as tangent vectors to the manifold of statistical \tc{distributions}, we relate the result to the choice of a coordinate patch that makes the Fisher-Rao metric Euclidean at the \tc{steady distribution}, realizing a sort of statistical equivalence principle. We \tc{then} classify nonequilibrium systems according to their spectrum, \tc{showing} that a degenerate Fisher matrix is the signature of the insurgence of a class of \tc{dynamical} phase transitions between nonequilibrium regimes, \tc{characterized by level crossing and power-law decay in time of suitable order parameters. An important consequence is that normal systems cannot manifest critical behavior. Finally, we study the Fisher matrix of systems with time-scale separation.}
}

\maketitle

{\color{\black}

\section{Introduction} The central tools of equilibrium statistical mechanics are the statistical ensembles (e.g. the micro-, macro- and grand-canonical). In the information-theoretic framework pioneered by Jaynes \cite{jaynes2}, ensembles can be obtained by maximizing the Gibbs-Shannon entropy at fixed average values of a set of extensive observables. The ensemble's dependence on the dual intensive variables (the Lagrange multipliers of the maximization procedure) characterizes response to perturbations, describing the thermodynamics of systems displaced from an equilibrium distribution to another equilibrium distribution. A static notion of ``thermodynamic distance'' between such equilibrium distributions has then been introduced in terms of the so-called Fisher-Rao metric, and its properties related to phase transitions \cite{equi,prop}.

Given the success of this approach, it is not surprising that many approaches to nonequilibrium systems are grounded on the same jargon and tools. Jaynes claimed that \textquotedblleft essentially all of the known results of Statistical Mechanics, equilibrium and nonequilibrium, are derivable consequences of this principle\textquotedblright, namely the maximum entropy principle  \cite{jaynes}. Unfortunately, the emphasis on the common features is often misleading. In fact, for nonequilibrium systems Jaynes's program has so far revealed to be inapplicable. Entropy maximization, the centrality and role of the ensemble, even the very existence of state functions, e.g. the energy, and a sensible definition of their intensive conjugates, e.g. the temperature, have to be renounced to. In particular, the statistical ensemble {\it per se} only contains a piece of information and  is insufficient to discriminate between equilibrium and nonequilibrium character. The phenomenology of nonequilibrium systems is significantly richer. In the dynamic nonequilibrium scenario, the system relaxes to a steady distribution where currents are constantly maintained within the system. This relaxation process is subtended by a complex interplay between decay modes.  Thus, the study of the modes of relaxation of a system is essential for the full characterization of nonequilibrium behavior.

It is then interesting to inquire whether the notion of statistical distance, and more generally the geometrical picture based on the Fisher-Rao metric, can be sensibly extended to nonequilibrium systems. In this paper we establish a connection between dynamic and thermodynamic aspects of nonequilibrium systems and geometrical properties of the Fisher-Rao measure of distance between probability distributions, when expressed in the preferred basis of decay modes of the system. The main message is that the interplay between decay modes signals the equilibrium/nonequilibrium nature of systems, the occurrence of dynamical phase transitions, the separation of time scales, and other interesting behavior. This information is encoded in the algebraic properties of the Fisher information matrix.}

We will model nonequilibrium systems by Markovian evolution of a probability distribution over a finite state space, decaying to a steady distribution. For such systems relative entropy with respect to the steady distribution plays the role of a Lyapunov functional. As was already known to Kullback \cite[Sec. 2.6]{kullback} (see Ref.\,\cite{baez1} for a simple review), relative entropy between nearby probability distributions also provides a definition for the Fisher's matrix $g^{\alpha\beta}$, which was reinterpreted by Rao as a metric $g$ on the manifold of statistical distributions \cite{fisher}. Already a standard tool in information theory and statistics \cite[Sec.1.4; Sec.1.5; Ch.14]{book}, the Fisher-Rao metric was rediscovered in equilibrium statistical mechanics \cite{equi}, and after a pioneering work by Obata \cite{obata} more recently it is drawing attention in nonequilibrium statistical mechanics \cite{crooks1}. \tc{The Fisher matrix for decay modes has been employed in a nonequilibrium context in Ref.\,\cite{polespo} to disprove a conjecture about convexity of the relative entropy}. It is also of great importance in quantum information  theory \cite{petz}, where singularities of $g^{\alpha\beta}$ have been shown to pinpoint quantum phase transitions \cite{zanardi}.

Analogously, we will show that the choice of a nonequilibrium generator picks out one preferred basis for the Fisher-Rao metric near the invariant distribution, in such a way that a diagonal Fisher matrix occurs for systems that commute with their time reversal \cite{maes}, which we refer to as \textit{$p$-normal}, or simply normal, while degeneracies are accompanied with the sort of critical behavior that occurs at nonequilibrium phase transitions ---but with no cogent need for a thermodynamic limit on the number of states. The results are based on a generalization of a well-known \cite[Sec. 5.7]{vankampen} orthonormality relation between decay eigenmodes of detailed-balanced systems to systems whose generators commute with their time reversal. To this class there belong equilibrium systems, with real decay spectrum, and a class of $p$-normal, truly nonequilibrium systems, with oscillatory behavior. \tc{As a consequence, for normal systems different decay modes contribute independently to the relative entropy towards the steady state, while for non-normal generators one has cross-mode contributions.} We finally linger on the geometric meaning of the results, speculating that decay modes can be seen as local reference frames for the Fisher-Rao metric, so that $p$-normal systems correspond to coordinate patches that make it Euclidean at one point of the manifold, thus realizing a sort of statistical equivalence principle analogous to the equivalence principle of General Relavity. \tc{Finally, we study the behavior of the Fisher matrix for systems that have a time scale separation between local and global relaxation.}
 
\section{Setup}

\tc{We consider a system with finite state space, with $n+1$ states labelled by roman indices $i,j,k$. On it, a normalized probability distribution $p^t = (p^t_i)_i$ undergoes Markovian evolution, $\tfrac{d}{dt}p^t  = L p^t$ with generator $L$ (we adopt the physicists' convention that the action of generators to the right governs the evolution of probability densities).} 
Conservation of probability requires the generator's columns to add to zero,
\be
L_{ij} = \left\{\ba{ll} w_{ij}, & i \neq j \\
- \sum_{\color{\black}k\neq i} w_{ki}, & i = j
\ea
\right.  .
\ee
The $w$'s are real positive transition rates, with units of an inverse time. 
Under mild assumptions on the connectedness of the state space and on the reversibility of paths, the Markovian dynamics has a unique steady distribution $p$, i.e. a null eigenvector of the generator, $Lp  = 0$, towards which any initial distribution $p_{\boldsymbol{\varepsilon}}$ tends at large times, $\lim_{t \to \infty} e^{tL} p_{\boldsymbol{\varepsilon}} = p.$
The propagator $\exp tL$ is a stochastic matrix with strictly positive entries.  By a standard argument in the theory of Markov chains, the Perron-Frobenius theorem can be applied to prove that the $n$ nonnull eigenvalues of $L$ have negative real part,
\be
L q^\alpha = \left(- 1/\tau_\alpha + i \omega_\alpha \right) q^\alpha, \qquad \alpha = 1,\ldots, n,
\ee
with $\tau_\alpha$ a positive characteristic decay time and $\omega_\alpha$ a frequency. \tc{Notice that, in the following, the greek index $\alpha = 1,\ldots, n$ labels decay modes relative to nonvanishing eigenvalues, while latin index $a = 0,\ldots,n$ also includes the eigenvector relative to eigenvalue zero, and we remind that $i$ spans the system's state space.} For the time being, we suppose the spectrum to be nondegenerate.  The decay eigenmodes  $q^\alpha$, plus the invariant distribution $p$, form a basis of eigenvectors of $L$. Propagating $p_{\boldsymbol{\varepsilon}} = \varepsilon_0 p  + \sum_\alpha \varepsilon_\alpha q^\alpha$ up to time $t$ yields
\be \exp (tL) p_{\boldsymbol{\varepsilon}}  = \varepsilon_0 p  + \sum_\alpha e^{-t/\tau_\alpha + i \omega_\alpha t}  \varepsilon_\alpha q^\alpha.\label{eq:timev} \ee
At late times decay modes are exponentially damped, so that by normalization $\varepsilon_0 = 1$, and consequently the eigenmodes' entries are seen to add to zero, $\sum_i q_i^\alpha = 0$.

\section{The Fisher matrix}

Following Andrieux \cite{andrieux}, we define the diagonal matrix
\be
\sqrt{p }= \mathrm{diag} \{\sqrt{p_1},\ldots,\sqrt{p_{n+1}}\}
\ee
and introduce the tilted generator
\be
H = \sqrt{p }^{\;-1} L \sqrt{p }. \label{eq:energy}
\ee
This transformation is a similarity of matrices, hence the spectrum of $H$ coincides with that of $L$, $H e^0 = 0$, $H e^\alpha = (-\tau_\alpha^{-1} + i \omega_\alpha) e^\alpha$. \tc{In particular eigenvector $e^0$ has entries $e^0_i = \sqrt{p _i}$. All other eigenvectors $e^\alpha$ are related to the decay modes via $e^\alpha_i =  q_i^\alpha/\sqrt{p _i}$}.

\tc{Let us for the moment consider a generator $L$ with real spectrum. We build the mutual superpositions of the eigenvectors of $H$ with respect to the Euclidean scalar product $(\cdot,\cdot)$}
\be
h^{ab} = (e^a, e^b). \label{eq:extmetric}
\ee
The left-hand side provides the definition of the \tc{$(n+1)\times (n+1)$} symmetric matrix $h$. Entries along the zero-th row yield the identities $h^{00} = \sum_i p _i = 1$ and $h^{0\alpha} = \sum_i q_i^\alpha = 0$. \tc{Then, in all generality we have}
\be
h = \left(\ba{cc} 1 & 0 \\ 0 & g \ea \right) \label{eq:nove}
\ee
\tc{where the entries of the $n\times n$ block $g$, with respect to the nonnull eigenvectors of $L$,  are given by}
\be
g^{\alpha\beta} = \sum_i \frac{q_i^{\alpha} q_i^{\beta}}{p_i }. \label{eq:result}
\ee
{\color{\black} We call matrix $g$ the {\it Fisher matrix}. We will comment on the meaning of the Fisher matrix in estimation theory, and its interpretation in geometrical terms, in Secs. \ref{dpt} and \ref{geometrical}. Let us further introduce  the {\it volume element} 
\be
\eta = \frac{\det g}{\prod_\alpha g^{\alpha\alpha}}
\ee
as an indicator of the degree of superposition of the eigenmodes. 
Since $g$ is a Gramian matrix, by the Hadamard inequality \cite{dragomir} $\eta$ ranges between $0$ (when two modes overlap) and $1$ (when all modes are independent), having the geometric meaning of volume of the parallelotope enclosed by the normalized vectors $e^\alpha/\|e^\alpha\|$.}

\tc{Let us now consider a generator with complex spectrum.} Since the entries of $L$ are real, its $2k\leq n$ complex eigenvectors come in complex conjugate pairs $(q^\kappa_{+})^\ast =q^\kappa_{-}$, \tc{where we let index $\kappa = 1,\ldots, k \leq n/2$ range over couples of complex conjugate eigenvalues, and index $\iota$ range over the remaining real eigenvalues.} 
\tc{For the definition of the Fisher matrix, the general idea is that we employ the real and imaginary parts of each complex eigenvector. The result is a {\it complexified} Fisher matrix in the form}
{\color{\black}
\be
\tilde{g} = \left(
\ba{ccc}
g_{\Re\Re} & g_{\Re\Im}  & g_{\Re} \\
{g_{\Re\Im}}^T & g_{\Im\Im} & g_{\Im} \\
{g_{\Re}}^T & {g_{\Im}}^T & g \ea \right)
\ee 
where $^T$ denotes transposition and the blocks have entries
\bea
{g_{\Re\Re}}^{\kappa\kappa'} = (\Re e^\kappa, \Re e^{\kappa'}), & \quad &
{g_{\Im\Im}}^{\kappa\kappa'} = (\Im e^\kappa, \Im e^{\kappa'}), \nonumber \\
{g_{\Re\Im}}^{\kappa\kappa'} = (\Im e^\kappa, \Re e^{\kappa'}), & \quad &
{g_{\Im}}^{\kappa\iota} = (\Im e^\kappa, e^{\iota}), \nonumber \\
{g_{\Re}}^{\kappa\iota} = (\Re e^\kappa, e^{\iota}), & \quad &
g^{\iota\iota'} = (e^\iota,e^{\iota'}).
\eea
}

\section{Detailed balance and normal systems}

{\color{\black}
Generator $L$ satisfies detailed balance when the steady currents $\j _{ij} = w_{ij} p_j  - w_{ji} p_i $ all vanish, there being no net exchange of information between states. In this case the steady distribution is said of equilibrium. Otherwise it is a nonequilibrium steady distribution. It is a simple fact that the generator is detailed balanced if and only if $H$ is symmetric. Then,} by the spectral theorem $H$ admits a set of orthonormal eigenvectors \tc{relative to real eigenvalues. One then obtains for the Fisher matrix}
\be g^{\alpha\beta}  = \delta^{\alpha\beta} \ee
\tc{and the volume element attains its maximum value $\eta = 1$.} We can then state that $g^{\alpha\beta}$ is diagonal in a basis of eigenmodes of a Markovian generator with real non-degenerate spectrum if and only if the latter is of equilibrium. \tc{In Sec. \ref{geometrical} we will interpret this fact as a sort of ``equivalence principle'' for Markov dynamics.}

Let us further define the \textit{time-reversal} (also known as $p$-dual)
Markovian generator
\be
\bar{L} = \sqrt{p}^{\;2} L^{T} \sqrt{p}^{\;-2},
\ee
which also has $p$ as its invariant distribution \cite{esposito}. Generator $\bar{L}$ inverts all of the steady currents $\bar{\j}_{ij} = - \j_{ji}$. \tc{Moreover, the eigenvalues of the time-reversal generator coincide with those of $L$, but the corresponding eigenvectors are mapped into the complex-conjugate of those of $L$, so that the oscillatory behavior inverts the frequency of decay modes while preserving the decaying character. For detailed-balanced systems, one has $L=\bar{L}$.}

The spectral theorem generalizes to normal matrices: it suffices (and is necessary) that $HH^\dagger=H^\dagger H$ to make $H$ unitarily diagonalisable, i.e. with orthonormal eigenvectors with respect to the Hermitian scalar product $( v,w)_{\mathbf{C}} = \sum_i v_i^\ast w_i$. 
Normality of $H$ translates in the commutation relation of $p$-normal generators:
\be [L,\bar{L}]=0 .\label{eq:commut} \ee
It is then simple to see that Eq.\,(\ref{eq:commut}) is a sufficient and necessary condition for the real and imaginary parts of $e_{\pm}^\kappa = \sqrt{p}^{\ -1} q_{\pm}^\kappa$ to satisfy
\be
{g_{\Re}}^{\kappa\kappa'} =\tfrac{1}{2} \, \delta^{\kappa\kappa'}  =  {g_{\Im}}^{\kappa\kappa'},
\ee
to be orthogonal among themselves, $(\Re e^\kappa, \Im e^{\kappa'})= 0$, and to the remaining $n-2k$ real eigenvectors $e^\iota$, which satisfy Eq.\,(\ref{eq:result}) on their own, $(e^\iota,e^{\iota'})= \delta^{\iota\iota'}$. \tc{Hence, $p$-normal generators are those for which the complexified Fisher matrix is diagonal.}

\section{Relative entropy asymptotics}

{\color{\black}

Let us introduce the relative entropy with respect to the steady distribution along a solution of the master equation
\be
S(p^t \, |\, p) = \sum_i p_i^t \ln( p_i^t /p_i ). \label{eq:relent}
\ee
In this section we relate the Fisher matrix to the behavior in time of the relative entropy with respect to the steady distribution when state $p^t$ is sufficiently close to the steady state.

To this purpose,} we distinguish $\lfloor n/2 \rfloor $ phases in the space of generators with non-degenerate spectra, parametrized by transition rates.
\vskip 0.2cm
(A) There are $n$ real negative eigenvalues.  Expanding Eq.\,(\ref{eq:relent}) to second order in $\boldsymbol{\varepsilon}^t$ near the steady distibution, with $p^t = \sqrt{p}\, (e^0 + \varepsilon^t_\alpha e^\alpha)$, we obtain
\be
S(p^t \, |\, p) ~\approx ~ \tfrac{1}{2} \sum_{\alpha,\beta} \varepsilon^t_\alpha \varepsilon^t_\beta (e^\alpha,e^\beta) ~ =~ \tfrac{1}{2}  \| \,  \bolde^t \, \|_{g}^2\; , \label{eq:fish1}
\ee
where first order contributions vanish. To second order, relative entropy is one-half the lenght of vector $\bolde^t$ with respect to the positive-definite metric $g^{\alpha\beta}$ defined in Eq.\,(\ref{eq:result}). Considering the explicit time evolution, Eq.\,(\ref{eq:timev}), with $\omega_\alpha=0$, we identify  $\varepsilon^t_\alpha = e^{-t/\tau_\alpha} \varepsilon_\alpha$. Nonequilibrium models with real spectrum then display superposition of modes with different decay times, affecting the late time behavior of relative entropy. This superposition disappears for equilibrium generators. In a way, $g^{\alpha\beta}$ measures the correlation between decay eigenmodes, with equilibrium modes being uncorrelated \tc{and each contributing an independent term to the relative entropy}.

\vskip 0.2cm
(B$^k$) There are  $n-2k$ real eigenvalues $-1/\tau_\iota$ and $k$ couples of complex conjugate eigenvalues $ - 1/\tau_\kappa \pm i \omega_\kappa$. In a basis of eigenmodes, we express the initial distribution as 
\begin{subequations} \label{eq:complex}
\bea
p_{\bolde} & = & \sqrt{p }\left[e^0 + \sum_\iota \varepsilon_\iota e^\iota +  \nicefrac{1}{2} \sum_{\kappa} \left( \varepsilon_\kappa^+ e^\kappa_+ +  \varepsilon_\kappa^- e^\kappa_-  \right) \right] \nonumber \\ & = & \sqrt{p}\left[ e^0 + \sum_\iota \varepsilon_\iota e^\iota + \sum_\kappa (\varepsilon_\kappa^1 e^\kappa_1 + \varepsilon_\kappa^2 e^\kappa_2) \right]. 
\eea
\end{subequations}
In the first line, the factor $\nicefrac{1}{2}$ is there for sake of convenience. Since $p_{\bolde}$ is real, one has $(\varepsilon_\kappa^-)^\ast = \varepsilon_\kappa^+$. In the second line, we employ $e^\kappa_1= \Re e^\kappa_+ $, $e^\kappa_2 = \Im e^\kappa_+$ and $\varepsilon_\kappa^1= \Re \varepsilon_\kappa^+$, $\varepsilon_\kappa^2 = \Im \varepsilon_\kappa^+$. Propagating the initial distribution up to time $t$ we obtain, in full extent, 
\begin{multline}
e^{tL} p_{\bolde} ~=~
\sqrt{p}~ \bigg\{  e^0 + \sum_\iota e^{-t/\tau_\iota} \varepsilon_\iota e^\iota + \\
 + \sum_\kappa e^{-t/\tau_\kappa}  \Big[\cos(\omega_\kappa t) \varepsilon_\kappa^1  e^\kappa_1 -  \sin(\omega_\kappa t) \varepsilon_\kappa^1 e^\kappa_2 \\
  +\sin(\omega_\kappa t)  \varepsilon_\kappa^2 e^\kappa_1 +\cos(\omega_\kappa t)  \varepsilon_\kappa^2 e^\kappa_2
\Big]~ \bigg\} \;. \label{eq:multi}
\end{multline}
We define the block-diagonal matrix 
\be
 \Omega^t = \left(
\ba{cc}
e^{-t/\tau_\kappa} R(\omega_\kappa t) & 0   \\
0  &  e^{-t/\tau_\iota}\ea \right),   \label{eq:omega}
\ee
which has $k$ copies of $R(\varphi)$, the matrix of planar rotations by an angle $\varphi$. We can then recast Eq.\,(\ref{eq:multi}) in short as 
$e^{tL} p_{\bolde}  = \sqrt{p} (e^0 +   \tilde{\boldsymbol{e}}^T \Omega^t \tilde{\bolde})$. While relative entropy does not look like a real bilinear form when expressed in terms of the complex vector $\boldsymbol{\varepsilon} = (\varepsilon_\kappa^+,\varepsilon_\kappa^-,\varepsilon_\iota)$, it is indeed a positive bilinear form of the complexified vector $\tilde{\boldsymbol{\varepsilon}} = (\varepsilon_\kappa^1 , \varepsilon_\kappa^2,\varepsilon_\iota)$:
\be
S(p^t | \ p) \approx \tfrac{1}{2} \| \, \Omega^t \tilde{\boldsymbol{\varepsilon}} \,  \|_{\tilde{g}}^2\; , \label{eq:rotation} 
\ee 
where the norm is to be calculated using the complexified metric. Eq.\,(\ref{eq:rotation}) displays mixing of decay and oscillatory times, which only disappears when  the generator commutes with its reversal and $\tilde{g}^{\alpha\beta}$ is diagonal, with $k$ blocks $\propto \mathbf{1}_{2 \times 2}$. In fact, when $L$ is  $p$-normal, from the Hermitian orthogonality relations
\be
(e_1 + ie_2,e_1+ie_2)_{\mathbb{C}} = 0 , \quad (e_1 - ie_2,e_1+ie_2)_{\mathbb{C}} = c 
\ee
there follow the real orthonormality relations
\be
(e_1, e_2) = 0, \quad (e_1,e_1) = (e_2,e_2) = \tfrac{c}{2}.
\ee
Whatever the normalization chosen for the complex eigenvectors, real and imaginary parts have the same normalization, so that $\tilde{g}$ consists of blocks of type $c \mathbf{1}_{2\times 2}$. It follows that for $p$-normal systems oscillatory late-time behaviour in the relative entropy, encoded in ${\Omega^t}^T \tilde{g} \Omega^t$, will disappear, since $R(\omega_\kappa t)^T \cdot c \mathbf{1}_{2\times 2} \cdot  R(\omega_\kappa t) = c \mathbf{1}_{2\times 2}$.

\section{Dynamical phase transitions\label{dpt}}

One last case is left out from \tc{the above} analysis:
\vskip 0.2cm
(C) The generator is defective, that is, degenerate eigenvalues lack a complete set of eigenvectors. \tc{In other words, there is a {\it level crossing} analogous to that appearing in the phenomenology of quantum phase transitions.}

\begin{figure*}[tb]
  \centering
 \def\svgwidth{300pt}
 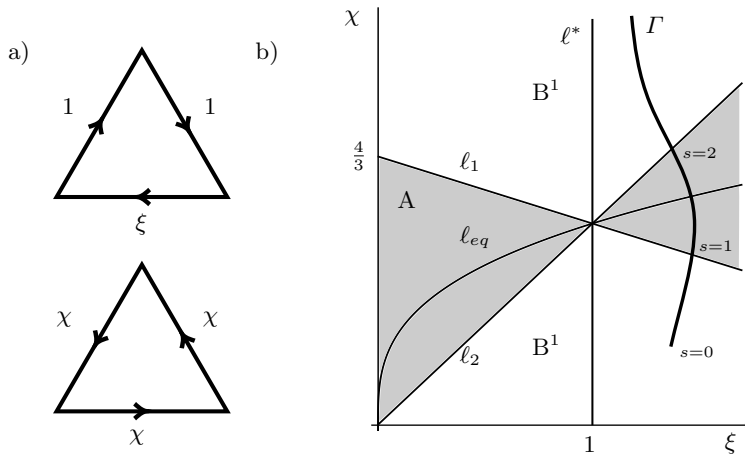
 \caption{\label{fig:phases}  a) Transition rates for the hopping biased particle with one clockwise perturbed rate. \tc{The arrows indicate the direction of the transition.} b)  The parameter space of $L(\xi,\chi)$: phases, critical and trivial curves, a path $\Gamma$.}
\end{figure*}

This case and the phenomenology so far analysed are better illustrated with the aid of an example, \tc{the full treatment of which can be found in Appendix \ref{app3}.} Consider the following generator, parametrized by positive rates $\xi, \chi$
\be
L(\xi,\chi) = \left( \ba{ccc}
- \xi-\chi & 1 & \chi  \\ 
\chi  & -1 - \chi & 1  \\
\xi & \chi & -1-\chi
\ea \right) .
\ee 
The dynamics generated by $L(\xi,\chi)$ is that of a hopping particle with a systematic bias in the counterclockwise direction, and one perturbed clockwise rate (see Fig.\ref{fig:phases}a).  Its phase space is depicted in Fig.\ref{fig:phases}b. \tc{The class of detailed balanced systems is found by applying Kolmogorov's criterion, stating that the ratio of the product of transition rates along a cycle, over the product of transition rates along the inverse cycle, should yield one, for all cycles. In this case there is only one cycle yielding $\xi/\chi^3 =1$,} which identifies the equilibrium \tc{curve} $\ell_{eq}$. The  model corresponding to $\chi = 1 =\xi$ is known as the unbiased hopping particle,
with twice degenerate eigenvalue $\lambda = 2$ affording a complete basis of eigenvectors. The space of parameters is partitioned into two phases of type A (in grey in Fig.\ref{fig:phases}a) and B$^1$ (in white), marked out by the critical lines $\ell_1: \chi+3\xi=4 $ and $\ell_2 : \xi=\chi$ \tc{that correspond to values of the parameters for which there only is one eigenvalue of algebraic multiplicity $2$ (see Appendix \ref{app3}).} For the first class of models, direct calculation of the eigenvectors shows that $g^{\alpha\beta}$ is diagonal only  along the equilibrium line. In phase B$^1$ one needs to turn to the complex components of the eigenmodes to be able to expand relative entropy as a positive bilinear form. Along  $\ell^\ast$ are the biased hopping particle models $L(1,\chi)$, which make the complexified matrix diagonal. Their reversal if found by inverting the bias in the clockwise direction, yielding $\bar{L} = L^T$.

With the exception of $L(1,1)$, a generator $L$ picked along the critical lines only  has one eigenvector $q$ relative to the degenerate eigenvalue $- \tau^{-1}$, \tc{i.e. the geometric multiplicity is $1$}. A generalized eigenvector $u$ shall then be introduced,  with $L u = - \tau^{-1} u + q$, carrying $L$ into Jordan's normal form. The time evolved $\exp(tL)p_{\bolde}$ is seen to acquire a term $\propto t e^{-t/\tau} q$ \cite{andrieux}.

Consider now a path $\Gamma = \{\xi(s),\chi(s)\}$ in parameter space, as depicted in Fig.\ref{fig:phases}. We first traverse the complex phase. At  $s=1$ we come upon an abrupt switch in the appropriate basis. Approaching the critical line $\ell_1$ from below, the imaginary part $\Im q_+(s)$ becomes smaller and smaller until it vanishes; from above, the modes $q_+(s)$ and  $q_-(s)$, respectively with higher and lower eigenvalue, tend to align. At the critical line the correlation matrix becomes degenerate and has a discontinuity
\be g^\ua = \left( \ba{cc}1 & 0 \\ 0 & 0 \ea \right) \to g^\da = \left( \ba{cc}1 & 1 \\ 1 & 1 \ea \right). \ee
An order parameter $f_s(\infty)$ ---of little physical meaning though---
can also be engineered. Consider two vectors $q^\perp_-(s)$ and $u^\perp$, respectively in the orthogonal complement of $q_-(s)$  and $u$. Projecting $\exp [tL(s)]p_{\bolde}-p$ along the two, and then taking the ratio, yields
\begin{subequations}
\bea
f_{2 > s >1}(t) &\propto& \big[1 + c e^{-t \left(\tau_-^{-1} - \tau_+^{-1} \right)}\big]^{-1},  \\
f_1(t) & \propto & 1 /(1+ c' t).
 \eea
 \end{subequations}
The exponential decay, which reaches an arbitrary nonzero value of $f_s(\infty)$, becomes a power-law at the critical line, with $f_s(\infty)= 0$. \tc{This is precisely the kind of behavior displayed by dynamical phase transitions, see for example the simple example} treated in \cite[Sec. 1.1]{marro}. This is also motivated by the equilibrium usage of the Fisher matrix, which coincides with the covariance matrix of the observable constraints $X^\alpha$ that define the ensemble $p \propto \exp - \beta_\alpha X^\alpha$. Crooks \cite{crooks1} commented that, as we vary the intensive parameters $\beta_\alpha$, correlations vary smoothly except at phase transitions, where divergencies occur. On this line,
further insight  might come from estimation theory. \tc{Let us introduce certain stochastic variables} $\hat{\boldsymbol{\varepsilon}}^i = (\varepsilon_1^i,\ldots, \varepsilon^i_n )$ \tc{whose purpose is to evaluate the composition of the decay modes, for example to estimate the time at which a sample of the system is drawn. We require this estimator to be unbiased, that is its} average  $\langle \hat{\boldsymbol{\varepsilon}} \rangle_{p_{\boldsymbol{\varepsilon}}} = \boldsymbol{\varepsilon}$ yields precisely the vector of parameters that specifies the initial distibution. The Cr\'amer-Rao inequality \cite{book}, establishes a lower bound on the covariance matrix $\langle (\hat{\bolde} - \bolde) (\hat{\bolde} - \bolde)  \rangle_{p_{\bolde}} \geq g^{-1}$, where $A \geq B$ means that $A-B$ is positive semidefinite.
Multiplying by $g$ and taking the trace we obtain
\be
\langle \| \, \hat{\bolde}- \bolde \, \|^2_{g(s)} \rangle_{p_{\bolde(s)}} ~\geq~ n, \quad s \neq 1,2. \label{eq:diver}
\ee
A degenerate metric admits non-null vectors of null norm. Hence in the limit $s \to 1$ certain correlations between unbiased estimators of the parameters diverge, since degeneracy of the bilinear form must be compensated by a divergence in order to verify Eq.\,(\ref{eq:diver}).

\tc{Thus we are able to relate the degeneracy of the Fisher matrix to a number of signatures of phase transitions: level crossing (i.e. overlapping eigenvalues), order parameters with polynomial decay in time, and divergent correlations of suitable observables. Notice that for this kind of systems the volume element $\eta = 0$. Hence, as an interesting consequence, one obtains that normal systems with  $\eta = 1$ (including detailed balanced systems) cannot display critical behavior, which is then an intrinsically nonequilibrium, non-normal behavior.} 

\section{Geometrical interpretation \label{geometrical}}

Indeed, a rich nonequilibrium phenomenology is marked by the peculiar representation of relative entropy near the steady distribution \tc{in terms of decay modes}, which can be interpreted as a metric on the manifold of statistical distributions $\mathcal{P}$.
Let us hint at its \tc{geometrical rationale, leaving further details to Appendix \ref{app1}}. 

With an information-theoretical attitude, one would like to employ relative entropy as a tool to compare probability distributions. However, relative entropy is not a good distance: it is not symmetrical, and the triangle inequality can be violated  \cite{baez1}, as one can split a path between two far-apart points into short segments whose relative entropies add up to a number smaller than $S(p' \vert p )$.
The way out of this puzzle is to stick to nearby distributions, as we did in Eq.\,(\ref{eq:fish1}), thus obtaining a local metric that measures the length of vectors $\varepsilon_\alpha q^\alpha$ living on the tangent space to $\mathcal{P}$ at $p$ (in this section the Einstein convention on index contraction is assumed). When moving to a different neighbourhood, one will shift the reference probability distribution to $p'$, and there define the metric in terms of $S(\,\cdot \, |\, p')$. If this procedure is carried on point-wise, one endows $\mathcal{P}$ with the Fisher-Rao metric. One can then assign coordinates $x^\alpha$ to neighbourhoods of the manifold; associated to such coordinates is a basis of preferred tangent vectors $\nicefrac{\partial}{\partial x^\alpha}$, which yield a matrix representative for the metric at each point of the neighbourhood. Notice that the Fisher-Rao metric is smoothly defined all over the manifold (except at boundaries and corners); it is its coordinatisation that might suffer from pathologies, as is the case for our critical systems.

But for $n=1$, it can be shown that $g$ has a non-null Riemann curvature:  while one can always choose a coordinate patch that trivializes  the metric at one given point, there is no such coordinate transformation which simultaneously makes $g$  diagonal all over a neighbourhood. A consequence of employing a nonflat metric is that the ``thermodnamic lenght'' of quasi-static processes, as discussed by Crooks \cite{crooks1}, depends on the path along the manifold.
Given the twice-contravariant transformation law for the metric, 
${g}^{\alpha'\beta'}= \Lambda_{\alpha}^{\alpha'} \Lambda_{\beta}^{\beta'}  g^{\alpha\beta}$, where  $\Lambda_\alpha^{\alpha'}(x') =\nicefrac{\partial x_\alpha}{\partial x'_{\alpha'}}$ is the inverse Jacobian of the coordinate transformation $x \to x'(x)$, after eqs.(\ref{eq:extmetric},\ref{eq:nove}) one realizes that the components of $e^\alpha_i$ can be interpreted as the Jacobian of an embedding patch, also called a \textit{frame}, which trivialises the metric at $p$. Hence, orthonormal frames are associated to $p$-normal systems; vice versa, two such systems with the same steady distribution yield different orthonormal frames, which are connected by a {\it gauge} transformation in the special group of orthogonal transformations.

This very mechanism lies at the heart of the Equivalence Principle of General Relativity. While gravity curves spacetime so as to prevent the definition of broad notions of ``parallelism'' and ``simultaneity'', one can always find  coordinates that make spacetime Minkowskian at one point, and gravity indiscernible from a fictitious force.  To a special observer, the frames' entries  provide an inertial frame of coordinate axis: in a very precise way they measure how much the orientation of these axis differs, up to Lorentz gauge transformations, from  ``bent'' coordinate axis. This discrepancy \textit{is} the gravitational field \cite[pp. 59-60]{rovelli}.

{\color{\black}
\section{\label{coarse}Coarse graining}

In this section we study the behavior of the Fisher matrix when the system has two different time scales. More precisely we consider a Markov generator in the form 
\be
L = \left(\ba{cccc} L^{(11)} & \epsilon L^{(12)} & \ldots & \epsilon L^{(1m)} \\
\epsilon L^{(21)} & L^{(22)} & \\
\vdots & & \ddots \\
\epsilon L^{(m1)} & & & L^{(mm)}
 \ea \right) 
\ee
where $L^{(ij)}$ is a block of dimension $(n_i+1) \times (n_j+1)$, with $\sum_{j=1}^m (n_j+1) = n+1$. The parameter $\epsilon$ is taken to be small. 
Such generators have been considered in the context of stochastic thermodynamics by Esposito \cite{espositocoarse}. It is intuitively clear that equilibration occurs faster in the subspaces with higher rates. An analogous situation is encountered in continuous-state space with overdamped Langevin equations, where momentum is the fast variable and position the slow variable.

Let 
\be
p = \left( \ba{c} p^{(1)} \\ p^{(2)} \\ \vdots \\ p^{(m)} \ea \right).
\ee
be the steady state of $L$, with the $p^{(i)}$ of dimension $n_i+1$. From the eigenvector equation we obtain $L^{(ii)} p^{(i)} = O(\epsilon)$. Then the block matrices $L^{(ii)}$ are to lowest order stochastic matrices, affording a normalized eigenvector and $n_i$ decay modes.

As regards the decay modes of $L$, one has $n_i$ internal decay modes concentrated in each block, corresponding to finite eigenvalues. In fact, assuming the {\it anszat}
\be
q_{in} = \left( \ba{c} q^{(1)} \\ \epsilon q^{(2)} \\ \vdots \\ \epsilon q^{(m)} \ea \right)
\ee
one can easily see that $q$ is a null-trace eigenvector of $L$ provided that $q^{(1)}$ is one of the decay modes of $L^{(11)}$, to order $\epsilon^2$. The eigenvalue equation also produces $m-1$ equations for the $q^{(2)} \ldots q^{(m)}$ that can be iteratively solved in terms of $q^{(1)}$. Similarly one can proceed for the other blocks.

Finally, there are $m-1$ external modes. Assuming the {\it anszat}
\be
q_{ex} = \left( \ba{c} \gamma_1 p^{(1)} \\ \gamma_2 p^{(2)} \\ \vdots \\ \gamma_m p^{(m)} \ea \right) + \epsilon \left( \ba{c} \eta^{(1)} \\ \eta^{(2)} \\ \vdots \\ \eta^{(m)} \ea \right),
\ee
in the eigenvalue equation one obtains all terms of order $\epsilon$, with a slow eigenvalue of order $\epsilon$ as well. In principle such equations can be solved to determine $\eta^{(m)}$ and the coefficients $\gamma_i$; notice that the vector $\gamma$ with entries $\gamma_i (1^T p^{(i)})$ has the meaning of a collective decay mode for the coarse-grained dynamics.

From the above structure of the eigenvectors it follows immediately that to lowest order the Fisher matrix attains a block-diagonal shape
\be
g = \left(\ba{ccccc} g_{in}^{(11)} \\ & \ddots \\ & & g_{in}^{(mm)} \\ & & & g_{ex} \ea\right) + O(\epsilon).
\ee
where $g_{in}^{(ii)}$ is the Fisher matrix of the internal decay modes in the $i$-th block, and $g_{ex}$ is the Fisher matrix of the $m-1$ external coarse-grained modes $\gamma$. What is remarkable and not obvious in the above expression is that not only fast decay modes corresponding to different blocks are orthogonal (as should be expected), but that also the external modes are orthogonal to all of the internal modes, to order $\epsilon$. This also implies that the internal and the external modes contribute independently to the relative entropy close to the steady distribution.
}

\section{Conclusions}

To conclude, the present paper establishes a  close connection between the Fisher-Rao metric and \tc{nonequilibrium properties of} Markovian generators. The choice of a generator induces a natural identification  of a positive semi-definite matrix representation \tc{via a preferred coordinatization of the manifold of statistical states near the steady distribution. The metric is trivial} for generators that commute with their time-reversal, including equilibrium systems. \tc{Critical behavior typical of} nonequilibrium phase transitions is induced by a degenerate Fisher matrix. \tc{Furthermore, the Fisher matrix has a peculiar behavior when the system presents separation of time scales. The shape of the Fisher matrix affects late-time behavior of the relative entropy with respect to the steady distribution}.

This study calls for a careful treatment of the algebraic varieties of critical and trivial loci in the space of generators, and for a thermodynamical characterization of the order parameters. Furthermore, normal systems seem to enjoy special properties, and deserve more in-depth study, in particular in relation to the fluctuations and large deviations of their microscopic jump trajectories. 
We notice at a passing glance that the ``curvature'' operator $[L,\bar{L}]\delta t^2$ retains a geometrical flavour, as it measures how circuitation along infinitesimal parallelograms fails to close, when we first evolve the initial distribution with $L$ up to time $\delta t$ and then run it back with $\bar{L}$, rather then first evolving it back to $- \delta t$ with $\bar{L}$ and then run it forward with $L$. Detailed fluctuation theorems revolve around the time-reversed generator \cite{esposito}.

Finally, the neat framework suggests to deepen the nonequilibrium characterization of geometric objects such as Christoffel coefficients, geodesic curves, intrinsic and extrinsic curvature. For example, an application of the algebraic properties here introduced to the poblem of convexity of the relative entropy along Markov dnamics has been given by the Author in Ref.\,\cite{polespo}.

To conclude with a motto, we might claim that equilibrium systems are to nonequilibrium thermodynamics  what inertial frames are to gravity.

\section*{Aknowledgments}

The author is grateful to M. Esposito and D. Andrieux for discussion, to J. Baez and D. Bianchini for discussion on the first draft. \tc{The research was partly supported by the National Research Fund Luxembourg in the frame of the AFR Postdoc Grant 5856127.}

\appendix

\section{\label{app1}On the Fisher-Rao metric}

The choice of a Markovian generator $L$ identifies a point on the manifold of statistical states and a set of decay modes. We first map the probability simpex into the surface of a sphere, then interpret the pushed-forward vectors' entries $e^a_i$ as the Jacobian of a second coordinate transformation, in such a way that an equilibrium generator corresponds to the choice of a coordinate patch which trivializes the Fisher-Rao metric at the invariant state.
Given an invariant state, there exists a whole $\mathrm{O}(n+1)$-orbit of  frame fields which trivialize the metric. Viceversa, two equilibrium generators with the same invariant state yield gauge-equivalent frames.

From a geometrical perspective ---see for example \cite{book}--- the two matrices $H$ and $-L$ are both representations of an operator $\mathbf{H}$  with eigenvectors 
\be \mathbf{e}^a = q^a_i \tfrac{\partial}{\partial p_i} = e^a_i \tfrac{\partial}{\partial z_i}.\ee
The peculiar notation employed for the basis vectors  denotes that  $\sqrt{p}$
might be seen as the inverse Jacobian of the coordinate transformation  $p_j \mapsto z_j(p) = 2 \sqrt{p_j}$, which maps the probability simplex $\{p_i \in [0,1]^{n+1} : \sum_j p_j = 1 \}$ into a portion of the hypersphere with square radius $\sum_i z_i^2 = 4$.  Both are embeddings of the abstract manifold $\mathcal{P}$ of probability distributions  into $\mathbb{R}^{n+1}$. Each such coordinatization (say, $x$)  of ``positions'' on $\mathcal{P}$ endows the $(n+1)$-dimensional vector space $V \cong \mathbb{R}^{n+1}$ of ``velocities'', attached to $p$, with a preferred  basis of directions $\nicefrac{\partial}{\partial x_i}$. Vectors $\mathbf{e}^\alpha$  span the $n$-dimensional  tangent space $T_{p} \mathcal{P} \subset V$, while $\mathbf{e}^0$ describes how a neighbourhood of $p$ sits in the embedding space.

\section{\label{app3}On the three-state system}

We work out in full extent the three-state model, with $x=\xi, y=\chi$.
Equilibrium holds when the Kolmogorov criterion is satisfied, that is then the ratio of products of clockwise over counterclockwise rates yields $1$. Its locus defines the equilibrium line $\ell_{eq} : x - y^3 = 0$. The characteristic polynomial of $L$ is
\be
\det (\lambda \mathbf{1} - L) = \lambda^3 + (x+2y+3) \lambda^2 + ( 3 y^2 +2x + 2y + xy + 1) \lambda,  \nonumber
\ee
with roots $\lambda = 0$ and $\lambda_{\pm} = - \left(x+3y+2 \pm \sqrt{\Delta} \right)/2$ where the discriminant $\Delta = (x-y)(x+3y-4)$ vanishes at the critical lines
$\ell_1 : x + 3y - 4 =0$ and $\ell_2 : x - y = 0$.
The unnormalized invariant state is
\be
p  =  (y^2+y+1, y^2+y+x , y^2 +xy+x).
\ee
We proceed analyzing the three cases of interest.

\vspace{0.2cm}
(A: $\Delta > 0$). The eigenvectors are
\be
q_{\pm} = \left(\ba{c}
4 y - (x+y\pm \sqrt{\Delta})^2\\
- 4 x + 2y \left(x + y  \pm\sqrt{\Delta} \right) \\
2 x^2 + 2 xy - 4y^2 \pm 2 x \sqrt{\Delta} 
\ea \right)  \nonumber
\ee
The off-diagonal element of the Fisher matrix reads
\bea
g^{+-} ~=~ \frac{ [4 y - (x+y + \sqrt{\Delta})^2]  [4 y - (x+y - \sqrt{\Delta})^2]}{y^2+y+1}  \nonumber  \\
+ \frac{ [- 4 x + 2y \left(x + y  +\sqrt{\Delta} \right) ]  [- 4 x + 2y \left(x + y - \sqrt{\Delta} \right) ]}{y^2+y+x}  \nonumber \\
+ \frac{ [2 x^2 + 2 xy - 4y^2 + 2 x \sqrt{\Delta} 
 ]  [2 x^2 + 2 xy - 4y^2 - 2 x \sqrt{\Delta} 
]}{y^2 +xy+x } . \nonumber
\eea
Imposing $g^{12}=0$, we find the solutions: $x - y^3 = 0$;  $x^2 - y^3 = 0$;  $1 + 2 x + 2 y + x y + 3 y^2 = 0$. The first is precisely the equilibrium line, which corroborates our conclusion. The third equation  has no physical solution. The second solution is more tricky, as we didn't expect its appearence. There the metric is \textit{not} diagonal, as $q_\pm$ are not the eigenvectors of $L(x,x^{2/3})$. In fact in this case
$q_+ = 0$, 
and there exists a correct complete basis of eigenvectors:
\be
\tilde{q}_{+} = \left(\ba{c}
x^{2/3} \\
-x^{2/3}-x\\
x
\ea \right) , \qquad
\tilde{q}_{-} = \left(\ba{c}
-x^{2/3}-x \\
x^{2/3}  \\
x
\ea \right) . \nonumber 
\ee

\vspace{0.2cm}
(B: $\Delta < 0$) Eigenvectors $q_{\pm}$ have real and imaginary parts given by
\be
\Re q_+ = \left(\ba{c}
2x - x^2 +y^2 -2xy\\
y^2 + xy-2x \\
x^2 + xy - 2y^2
\ea \right),\quad
\Im q_+ = \sqrt{|\Delta|} \left(\ba{c}
-x-y \\
y \\
x
\ea \right) \nonumber
\ee 
The off-diagonal element of the complexified Fisher matrix reads
\bea
\tilde{g}^{12} &=& \frac{(x-1)}{p_1 p_2 p_3} \left( x^4 + 2 x^2 y + 4 x^3 y + x^4 y + 6 x^2 y^2 + 7 x^3 y^2 + \right. \nonumber \\
& & \left. 3 x y^3 + 13 x^2 y^3 + x^3 y^3 + 6 x y^4 + 5 x^2 y^4 - y^5 + 
   6 x y^5 \right) . \nonumber
\eea
It vanishes on the line $\ell^\ast: x=1$, but there is another class of solutions in the first quadrant. It can be shown that those solutions correspond, as above, to systems for which $q^\pm$ are not the correct eigenvectors. The time-reversal operator is
\be \bar{L}(x,y)
= \left( \ba{ccc}
- x - y &  \frac{y+y^2+y^3}{x+y+y^2}  & \frac{x+xy+xy^2}{x+xy+y^2} \\ 
\frac{x+y+y^2}{1+y+y^2} & -1-y & \frac{xy+y^2+y^3}{x+xy+y^2} \\
\frac{xy+xy^2+y^3}{1+y+y^2} & \frac{x+xy+y^2}{x+y+y^2} & -1-y 
\ea \right).
\ee 
Setting the commutator with $[\bar{L},L]$ to zero, yields
\be
 [\bar{L},L] = 
\frac{(x-1)(y^3-x)}{p_1\cdot p_2  \cdot p_3 } M = 0 \label{eq:system}
\ee
with
\bea
M_{11} & = & -y (x + 2 x y + 2 y^2 + y^3) \nonumber \\
M_{12} & = & (1 + y + y^2) (x + x y + 2 y^2) \nonumber \\
M_{13} & = & -(x + y^2) (1 + y + y^2) \nonumber \\
\ldots & & \nonumber
\eea
We don't need to specify more entries of $M$, as system (\ref{eq:system}) has as its unique solutions $x=y^3$ (the equilibrium line) and $\ell^\ast: x= 1$, as expected.

\vspace{0.2cm}
(C: $\Delta = 0$). Notice that on the critical lines $q_+ = q_-$ and $\Im q_+ = 0$. Looking for eigenvectors, on $\ell_2$ we obtain 
\be
\lambda \mathbf{1} - L(x,x) = -
\left( \ba{ccc}
1 & 1 & x \\ x & x & 1 \\ x & x & x
\ea \right),
\ee
whose kernel is 1-dimensional, unless $x=1$. Similar conclusions can be drawn on $\ell_1$.

\end{document}